\documentclass{article}
\usepackage[preprint]{spconfa4}
\usepackage{amsmath,amssymb,graphicx}
\usepackage{multirow}
\usepackage{booktabs}
\usepackage[ancient]{flushend}
\usepackage[hidelinks]{hyperref}
\usepackage[sort]{cite}

\title{AMBISEP: AMBISONIC-TO-AMBISONIC REVERBERANT SPEECH SEPARATION USING TRANSFORMER NETWORKS}
\name{Adrian Herzog, Srikanth Raj Chetupalli and Emanu\"{e}l A. P. Habets}
\address{International Audio Laboratories Erlangen\sthanks{A joint institution of the Friedrich-Alexander-Universit\"{a}t Erlangen-N\"{u}rnberg (FAU) and Fraunhofer IIS, Germany.}, Am Wolfsmantel 33, 91058 Erlangen, Germany}
\begin{document}
\ninept
\maketitle
\begin{abstract}
Consider a multichannel Ambisonic recording containing a mixture of several reverberant speech signals. Retreiving the reverberant Ambisonic signals corresponding to the individual speech sources blindly from the mixture is a challenging task as it requires to estimate multiple signal channels for each source.
In this work, we propose AmbiSep, a deep neural network-based plane-wave domain masking approach to solve this task. The masking network uses learned feature representations and transformers in a triple-path processing configuration.
We train and evaluate the proposed network architecture on a spatialized WSJ0-2mix dataset, and show that the method achieves a multichannel scale-invariant signal-to-distortion ratio improvement of 17.7 dB on the blind test set, while preserving the spatial characteristics of the separated sounds.
\end{abstract}
\begin{keywords}
Ambisonics, speech separation, transformer networks, reverberation
\end{keywords}
\section{Introduction}
\label{sec:intro}
Sound field recording using spherical microphone arrays (SMAs) is becoming more and more popular due to the emerging trend of immersive media applications such as, e.g., virtual reality~\cite{Zotter2019}.
The recorded scene, often, contains a mixture of sounds from different sources, ambience or noise. For spatial audio reproduction or object-based audio rendering, it is desirable to process these sound components individually, for example, to attenuate an undesired component while preserving the spatial characteristics of the scene. In the context of SMA processing~\cite{Rafaely2015,Jarrett2017}, the microphone signals are commonly converted to a full-sphere surround sound format, namely, Ambisonics prior to further processing. Ambisonics represents the sound field via different directional components, it is independent of the specific microphone array properties and the spatial resolution can be controlled via the Ambisonic order, which makes processing in the Ambisonic domain suitable for spatial audio analysis and reproduction. 

In recent years, several works were dedicated to spatial parameter estimation and source separation from Ambisonic mixtures~\cite{Teutsch2008,Khaykin2009,Epain2010,Epain2016,Fahim2017,Nikunen2018,Perotin2018,Perotin2019}, but only a few publications investigated the estimation of the Ambisonic signals of the individual sound components~\cite{Borrelli2018,Hafsati2019,Herzog2020}. 
While Ambisonic-to-Ambisonic noise reduction was considered in~\cite{Borrelli2018,Herzog2020}, Ambisonic-to-Ambisonic signal separation was investigated in~\cite{Hafsati2019} using the flexible audio source separation framework~\cite{Ozerov2012}. 
However, to the best knowledge of the authors, no deep learning-based method was proposed for Ambisonic-to-Ambisonic speech separation in the literature so far.

Remarkable advances have been obtained in recent years for the task of monaural, speaker independent, speech separation in noiseless conditions using deep neural networks~\cite{Wang2018,Luo2019,Luo2020,Subakan2021}.
In particular, masking in the learned time-feature domain~\cite{Luo2019} combined with long- and short-time modeling for long sequences~\cite{Luo2020} and transformer networks~\cite{Subakan2021} has achieved state-of-the-art separation performance.
In contrast to traditional monaural source separation, which estimates a monaural signal for each source, Ambisonic-to-Ambisonic separation requires the estimation of a multichannel Ambisonic signal for each component while preserving the spatial characteristics of the component.

In this work, we develop a deep learning method for Ambisonic-to-Ambisonic speech separation from a noise-less but reverberant Ambisonic mixture of two speech signals. We propose plane-wave domain masking using a transformer network, organized in a triple-path configuration to model the inter-channel, as well as long- and short-time relationships. The architecture is inspired by the monaural separation network in~\cite{Subakan2021} and the multichannel-to-multichannel speech enhancement architecture in~\cite{Pandey2021}. We evaluate the proposed method using the WSJ0-2mix speech dataset~\cite{Hershey2016}, spatialized using simulated Ambisonic room impulse responses for first-order Ambisonics. We quantify the signal- and spatial distortions using the multichannel signal-to-distortion ratio and the image-to-spatial-distortion ratio~\cite{Vincent2007}.
Illustrative audio examples, binaurally rendered, are available online\footnote{\url{https://www.audiolabs-erlangen.de/resources/2022-IWAENC-AmbiSep}}.

\section{Signal Model and Problem Formulation}
\label{sec:sigmod}
Consider a sound scene containing $J$ speech sources in a reverberant enclosure, recorded using a SMA with $M$ microphones. Let $\mathbf{x}^\mathrm{amb}$ be the $L$'th-order Ambisonic signal vector obtained from the SMA recording, viz.,
\begin{equation}
\mathbf{x}^\mathrm{amb}(n) = [\chi_{00}(n), \chi_{1-1}(n), \chi_{10}(n), ..., \chi_{LL}(n)]^T
\end{equation}
where $n = 0, ..., N-1$ denotes the discrete time sample, $(\cdot)^T$ the transpose operation and $\chi_{lm}(n)$ the Ambisonic signal of order $l=0,...,L$ and mode $m=-l,...,l$. Note, that the dimension of $\mathbf{x}^\mathrm{amb}(n)$ is $(L+1)^2$.
Different Ambisonic formats exist. In this work, we use, without loss of generality, the AmbiX format~\cite{Nachbar2011}.

Let $\mathbf{c}^\mathrm{amb}_j$ denote the Ambisonic signal vector corresponding to the $j$'th speech source. Considering a noiseless scenario, the Ambisonic mixture $\mathbf{x}^\mathrm{amb}$ can be modeled as,
\begin{equation}
\mathbf{x}^\mathrm{amb}(n) = \sum\nolimits_{j=1}^J \mathbf{c}^\mathrm{amb}_j(n) ~.
\end{equation}
The task considered in this work is to estimate the Ambisonic speech components $\mathbf{c}^\mathrm{amb}_1, ..., \mathbf{c}^\mathrm{amb}_J$ blindly from the Ambisonic mixture $\mathbf{x}^\mathrm{amb}$.
We note that the speech components include reverberation, unlike most monaural speech separation approaches in which the reverberation is considered as undesired.

\section{Proposed Method}
\label{sec:proposed}

\subsection{Overview}
We consider masking in the learned time-feature (TF) domain, popularized by~\cite{Luo2019} and successfully employed in state-of-the-art single-channel source separation approaches~\cite{Luo2020,Subakan2021}.
Therein, a TF encoder converts the time-domain signal into a non-negative TF representation, a mask is learned in the TF domain for each source and the time-domain source signals are reconstructed from the respective masked TF representations.

\begin{figure}[t]
\centering\includegraphics[width=0.48\textwidth]{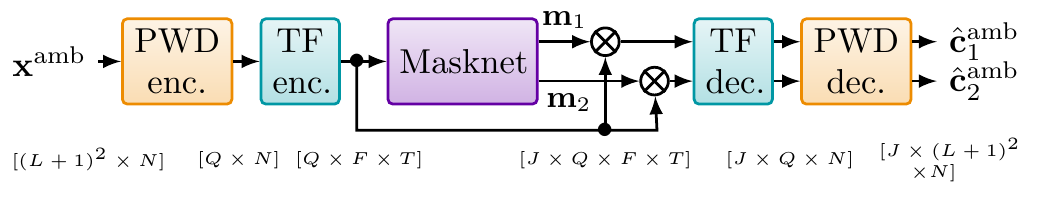}
\caption{High-level structure of the proposed Ambisonic separation method, shown for $J=2$.}\label{fig:highlevel}
\end{figure}

In this work, we propose a multichannel extension of the TF masking approach for Ambisonic-to-Ambisonic separation, denoted by AmbiSep. 
The most straight-forward approach would be to estimate TF masks for each Ambisonic channel. However, in the context of Ambisonic-to-Ambisonic noise reduction, it was shown that the plane-wave domain (PWD) masking approach from~\cite{Herzog2019a,Herzog2020} can result in a higher desired-signal quality~\cite{Lugasi2020}.
Therefore, we first transform the Ambisonic mixture $\mathbf{x}^\mathrm{amb}$ to the PWD via the PWD encoding matrix $\mathbf{Y}$, discussed in Sec.~\ref{suse:pwd}. Next, we transform each PWD channel to the TF domain using the encoding operator $\mathcal{E}$ and estimate mask vectors $\mathbf{m}_j(f,t)$ containing a TF mask per PWD direction for each speech component $j$, feature $f$ and time-frame $t$. After multiplying each PWD channel with the corresponding masks, we transform the signals back to the time domain using the TF decoding operator $\mathcal{D}$, which is not necessarily the inverse of $\mathcal{E}$, and to the Ambisonic domain using the pseudo-inverse PWD transformation matrix $\mathbf{Y}^\dagger$, i.e.,
\begin{equation}
\hat{\mathbf{c}}^\mathrm{amb}_j = \mathbf{Y}^\dagger \, \mathcal{D}\{ \mathbf{m}_j \odot \, \mathcal{E}\{\mathbf{Y} \, \mathbf{x}^\mathrm{amb}\} \}
\end{equation}
for $j=1,...,J$, where discrete-time and TF indices have been omitted for brevity and $\odot$ denotes the element-wise multiplication. The masks are estimated via the multichannel masking network (masknet) described in detail in Sec.~\ref{suse:masknet}. 
An overview of the proposed method is shown in Fig.~\ref{fig:highlevel}.

\subsection{Plane-Wave Domain Encoder and Decoder}\label{suse:pwd}
The Ambisonic signal vector $\mathbf{x}^\mathrm{amb}$ is projected to $Q=(L+1)^2$ almost uniformly distributed plane-wave directions $\Omega_1,...,\Omega_Q$ using maximum-directivity beamformers towards these directions. For SN3D-normalized Ambisonics, used by the AmbiX format, the maximum-directivity beamforming weights equal $\sqrt{2l+1}\,Y_{lm}(\Omega_q)$, where $Y_{lm}(\Omega_q)$ denotes the real spherical harmonic of order $l$ and mode $m$ at direction $\Omega_q$~\cite{Nachbar2011,Rafaely2015}. Therefore, the elements of the $Q \times (L+1)^2$-dimensional PWD encoding matrix $\mathbf{Y}$ become
\begin{equation}
[\mathbf{Y}]_{q,lm} = \sqrt{2l+1}\,Y_{lm}(\Omega_q)
\end{equation}
for $q=1,...,Q$, $l=0,...,L$ and $m=-l,...l$.
The PWD encoded mixture signal is denoted by $\mathbf{x}^\mathrm{pwd} = \mathbf{Y}\,\mathbf{x}^\mathrm{amb}$ in the following.
The PWD decoder applies the pseudo-inverse matrix $\mathbf{Y}^\dagger$ to the separated PWD outputs.

\subsection{Time-Feature Encoder and Decoder}
As in~\cite{Luo2019,Luo2020,Subakan2021}, we use a learnable TF encoder $\mathcal{E}$ and decoder $\mathcal{D}$. The encoded signal is computed via,
\begin{equation}
\mathcal{E}\{\mathbf{x}^\mathrm{pwd}\}(f,t) = \mathrm{ReLU}\{ \mathrm{conv1d}\{\mathbf{x}^\mathrm{pwd}\} \}(f, t) ~,
\end{equation}
where $\mathrm{conv1d}$ denotes a convolutional layer along the time dimension with $F$ filters of length $W$ and stride $H$. These parameters are analogous to the number of frequencies, the window length and the hopsize of a time-frequency transformation, respectively. The rectified linear unit ($\mathrm{ReLU}$) is applied after the convolution to yield a positive feature representation for the masknet.

The decoder consists of a transposed 1D convolutional layer with the same number of filters, length and stride parameters as the encoder. The same encoder and decoder pair is applied to all the PWD channels.

\subsection{Masknet}\label{suse:masknet}
\begin{figure}
\includegraphics[width=0.48\textwidth]{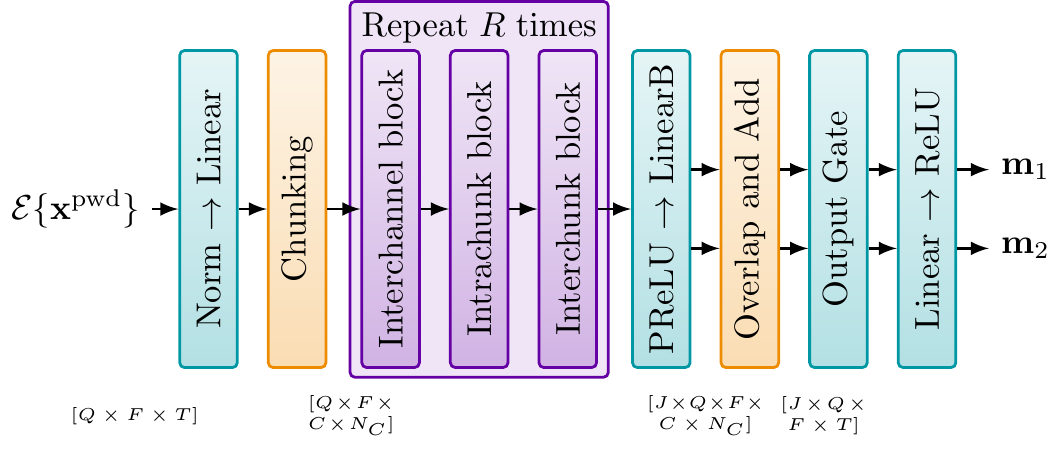}
\caption{Triple-path transformer masknet.}\label{fig:triplepath}
\end{figure}
The masknet follows the long- and short-time modeling approach for long-sequence modeling~\cite{Luo2020} using transformer blocks similar to the SepFormer~\cite{Subakan2021}. To handle the multichannel input, additional inter-channel transformer layers are incorporated, analogous to the triple-path approach proposed in~\cite{Pandey2021}.
The proposed masknet architecture is shown in Fig.~\ref{fig:triplepath}, where learnable blocks are colored turquoise or purple and purple blocks contain transformer layers.

The TF encoded PWD mixture of dimension $Q \times F \times T$, where $T$ denotes the number of time frames, is layer normalized per PWD channel followed by a linear layer along the feature dimension. The time dimension is then chopped into $N_C$ overlapping chunks of length $C$ with $50\%$ overlap. The chunked input representation is fed into the triple-path transformer network consisting of a sequence of inter-channel, intra-chunk and inter-chunk transformer blocks. This sequence is repeated $R$ times. Each block permutes the input tensor according to Tab.~\ref{tab:transformerdims} as in~\cite{Pandey2021} and applies $N_L$ transformer layers using the SepFormer architecture~\cite{Subakan2021} with a dimension of $N_\mathrm{FF}$ for the positional feed-forward networks. We refer the reader to~\cite{Subakan2021} for a detailed description of the transformer layers.
\begin{table}
\caption{Dimensions of transformer blocks.}\label{tab:transformerdims}
\begin{center}
\setlength\tabcolsep{5pt}
\begin{tabular}{lcccc}
\toprule
~ & Batch & Sequence & Feature & Tensor shape \\
\midrule
Interchannel & $CN_C$ & $Q$ & $F$ & $CN_C \times Q \times F$ \\
Intrachunk & $QN_C$ & $C$ & $F$ & $QN_C \times C \times F$ \\
Interchunk & $QC$ & $N_C$ & $F$ & $QC \times N_C \times F$ \\
\bottomrule
\end{tabular}
\end{center}
\end{table}

After the triple-path transformer network, a $\mathrm{PReLU}$ activation and an affine linear layer, i.e., linear plus bias (LinearB), are applied along the feature dimension with an output feature dimension of $J \cdot F$. The output is reshaped to $J \times Q \times F \times C \times N_C$ and the overlap and add method is used to de-chunk the signal. The respective output of dimension $J \times Q \times F \times T$ is fed to the output gate consisting of two paths. The first path transforms the input via an affine linear layer followed by a $\mathrm{tanh}$ activation and the second path via a different affine linear layer followed by a sigmoid activation. All transformations are applied along the feature dimension and the respective outputs are multiplied point-wise. Note, that this block is taken from the SpeechBrain implementation of the SepFormer and differs from the SepFormer architecture described in~\cite{Subakan2021}. Finally, the signal is fed into a linear layer followed by a $\mathrm{ReLU}$ activation to compute $Q$ masks for each of the $J$ sources.

\section{Baselines}
\label{sec:baselines}
As no deep learning-based Ambisonic-to-Ambisonic speech separation method has been published sofar, we define the following three baseline systems. 

\begin{figure}[t]
\centering\includegraphics[width=0.42\textwidth]{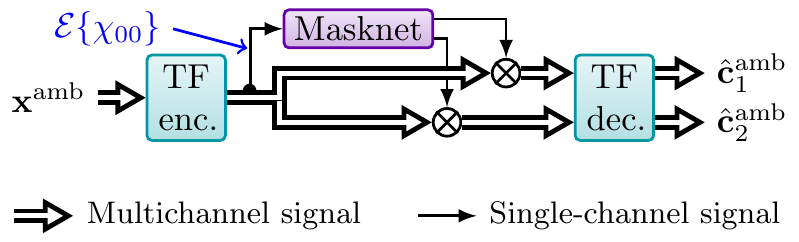}
\caption{Omni-SF baseline architecture.}\label{fig:omnisf}
\end{figure}

The first baseline, denoted by Omni-SF, computes a global TF mask $M_j(f,t)$ that is applied to all Ambisonic channels for each of the $J$ speech components, i.e.,
\begin{equation}\label{eq:globalmasks}
\hat{\mathbf{c}}^\mathrm{amb}_j = \mathcal{D}\{M_j \, \mathcal{E}\{\mathbf{x}^\mathrm{amb} \} \}
\end{equation}
for $j=1,...,J$. We use the SepFormer architecture on the omnidirectional (zeroth-order) input channel to compute the masks. The SepFormer model is trained using the globally-masked Ambisonic signals and the multichannel cost function discussed in Sec.~\ref{suse:training}. The Omni-SF architecture is shown in Fig.~\ref{fig:omnisf}

\begin{figure}[t]
\centering\includegraphics[width=0.42\textwidth]{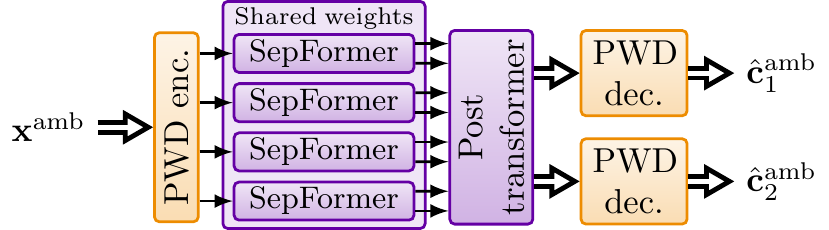}
\caption{PWD-SF baseline architecture.}\label{fig:pwdsf}
\end{figure}

For the second baseline, we compute masks for each PWD channel separately using the SepFormer architecture.
The estimated masks for the different speech components can be permuted across the different PWD channels.
To resolve this permutation problem, a single transformer layer (post transformer) is added before the PWD decoder. The time dimension of the input for the post transformer is chopped into $N_P = N / F$ non-overlapping blocks of length $F$ resulting in a tensor of shape $J \times Q \times F \times N_P$. The post transformer takes $N_P$ as the batch dimension, $J \cdot Q$ as the sequence length and $F$ as the feature dimension. The output is reshaped to $J \times Q \times N$. This baseline is denoted by PWD-SF in the following. The PWD-SF architecture is shown in Fig.~\ref{fig:pwdsf}.
Note that, in contrast to the proposed triple-path architecture, the PWD-SF cannot utilize spatial information before the final transformer layer.

For the third baseline, we use the system shown in Fig.~\ref{fig:highlevel} with oracle Wiener masks in the learned TF domain. The oracle masks are computed as,
\begin{equation}\label{eq:oraclemasks}
\mathbf{m}_j(f,t) = |\mathbf{c}_j^\mathrm{pwd}(f,t)|^2 \oslash \sum\nolimits_{j'=1}^J |\mathbf{c}_{j'}^\mathrm{pwd}(f,t)|^2
\end{equation}
for $j=1,...,J$, where $|\cdot |^2$ denotes the element-wise magnitude-squared, $\oslash$ the element-wise division and $\mathbf{c}_j^\mathrm{pwd}(f,t)$ the TF representation of $\mathbf{c}_j^\mathrm{pwd}(n) = \mathbf{Y}\,\mathbf{c}_j^\mathrm{amb}(n)$.

\section{Evaluation}
\label{sec:eval}

\subsection{Dataset}
We used the WSJ0-2mix dataset~\cite{Hershey2016} with 16 kHz sampling rate for the speech sources. The dataset contains 20k training samples, 5k validation samples and 3k test samples. Each sample comprises two English speech signals. All signals were limited to a maximum length of 5 seconds.

The reverberant Ambisonic speech mixtures were generated using simulated Ambisonic room impulse responses (RIRs).
We simulated SMA RIRs (SRIRs) using the SMA impulse response generator~\cite{Jarrett2012b} for a rigid SMA with 32 almost uniformly distributed microphones and a radius of $r=0.042$ m. The virtual SMA was positioned randomly in different simulated shoebox-shaped rooms with $T_{60}$ reverberation times ranging from $0.2 - 0.5 $ seconds. Different SRIR sets for the training, validation and test datasets were generated using 20, 5 and 3 different rooms, respectively, with 46 different SRIRs per room. The room dimensions $(L_x,L_y,L_z)$ were randomly sampled from $L_x, L_y \in [5\,\mathrm{m}, 10\,\mathrm{m}]$, $L_z \in [3\,\mathrm{m}, 4\,\mathrm{m}]$ and a minimum wall distance of $1\,\mathrm{m}$ was used for the SMA position. The speech sources were randomly positioned inside the room at distances $0.5 - 1.5\,\mathrm{m}$ from the array center and with a minimum wall distance of $0.5\,\mathrm{m}$.
The SRIRs were simulated with 16 kHz sampling frequency, an internal maximum spherical harmonic order of 20 and a length of $T_{60} / 2$.

We transformed the SRIRs to the Ambisonic domain using the discrete spherical harmonic transform with least-squares quadrature weights~\cite{Rafaely2015} and Tikhonov-regularized radial equalization filters~\cite{Politis2017} with a regulization factor of $\lambda_\mathrm{reg} = 10^{-3}$.
The SRIR dataset was generated with an Ambisonic order of $L=3$, but we used only the zeroth- and first-order channels for this work.

First-order Ambisonic mixtures were generated by randomly selecting a room and two SRIRs therein, convolving the speech sources with the SRIRs and mixing the respective components such that the power ratio of their zeroth-order channels matches the power ratio of the source signals.
The Ambisonic mixtures were generated dynamically with a different source--SRIR association at each training epoch for a given pair of sources in the original WSJ0-2mix dataset. This dynamic mixture creation was found to help with training compared to a fixed source--SRIR association for all the training epochs.

\subsection{Training and Parameters}\label{suse:training}
The models were trained using the Adam optimizer with a learning rate of $1.5 \cdot 10^{-4}$ for 170 epochs. The learning rate was decreased by a factor of $0.5$ after epoch 65 if the mean validation loss did not decrease for three consecutive epochs. For the loss function, we used the negative of the multichannel scale-invariant signal-to-distortion ratio (SI-SDR), viz.,
\begin{equation}\label{eq:multichannelsisdr}
\text{SI-SDR} = \frac{1}{J}\sum\limits_{j=1}^J 10\log_{10}\left( \frac{\sum_n \| \alpha_j\mathbf{c}_j(n)\|^2}{\sum_n \| \hat{\mathbf{c}}_{p(j)}(n) - \alpha_j\mathbf{c}_j(n)\|^2} \right) ~,
\end{equation}
where $\alpha_j = \mathrm{argmin}_\alpha \sum_n\| \hat{\mathbf{c}}_{p(j)}(n) - \alpha\,\mathbf{c}_j(n)\|^2$, $\|\cdot\|$ denotes the Euclidean norm and the correct target -- estimate permutation $p(j)$ for $j=1,...,J$ was determined using the utterance-level permutation-invariant training method~\cite{Kolbaek2017}. 
Note that a common scale $\alpha_j$ was chosen for all channels to enforce preservation of the spatial properties of the components.
Gradient clipping with a maximum 2-norm of 5 was applied during training.

The common model parameters are shown in Tab.~\ref{tab:params}.
We used the same number of transformer layers $N_L$ per block, i.e., $N_\mathrm{channel} = N_\mathrm{intra} = N_\mathrm{inter} = N_L$ and $R$ repeats for the triple-path transformer models.
\begin{table}
\caption{Model parameters.}\label{tab:params}
\begin{center}
\begin{tabular}{lll}
\toprule
Symbol & Description & Value \\
\midrule
$L$ & Ambisonic order & 1 \\
$Q$ & Number of PWD directions & 4 \\
$F$ & Number of TF encoding filters & 256 \\
$W$ & Encoding filter length & 32 \\
$H$ & Hopsize/stride of TF encoder & 16 \\
$C$ & Chunk size & 250 \\
$N_\mathrm{FF}$ & Feed-forward dim. of tranformers & 1024 \\
$N_\mathrm{heads}$ & Number of attention heads & 8 \\
$N_L$ & number of layers per transformer block & 1--8 \\
$R$ & number of repeats for transformer network & 1--8 \\
\bottomrule
\end{tabular}
\end{center}
\end{table}
For the oracle Wiener masking method, we replaced the masknet by the oracle Wiener TF masks~(\ref{eq:oraclemasks}) and trained the TF encoder and decoder for 50 epochs.

\subsection{Performance Metrics}
We used the SI-SDR improvement (SI-SDRi) and the scale-invariant version of the image-to-spatial-distortion ratio~\cite{Vincent2007} (SI-ISR) as performance metrics. The first metric measures spectro-temporal and spatial distortions, while the latter measures spatial distortions only. 

\subsection[Results]{Results}
\begin{table}
\caption{Test results for the investigated models.}\label{tab:results}
\begin{center}
\setlength\tabcolsep{3pt}
\begin{tabular}{l|lccccc}
\toprule
 & Method & $N_L$ & $R$ & SI-SDRi [dB] & SI-ISR [dB] & $\#$P \\
\midrule
\multirow{10}{*}{\rotatebox{90}{Proposed AmbiSep}} & \multirow{8}{*}{\parbox{5em}{\raggedright{iChan-First\\ and\\ with PWD}}} & $6$ & $1$ & $16.3 \pm 3.2$ & $22.9 \pm 3.4$ & $14.6\,$M  \\
 & & $3$ & $2$ & $17.2 \pm 3.3$ & $23.7 \pm 3.5$ & $14.6\,$M  \\
 & & $2$ & $3$ & $17.2 \pm 3.3$ &$23.8 \pm 3.5$ & $14.6\,$M  \\
 & & $1$ & $6$ & $17.3 \pm 3.3$ & $23.8 \pm 3.5$ & $14.6\,$M  \\
 & & $8$ & $1$ & $16.7 \pm 3.5$ & $23.4 \pm 3.7$ & $19.4\,$M  \\
 & & $4$ & $2$ & $17.6 \pm 3.4$ & $\mathbf{24.3 \pm 3.5}$ & $19.4\,$M  \\
 & & $2$ & $4$ & $17.5 \pm 3.2$ & $24.1 \pm 3.4$ & $19.4\,$M  \\
 & & $1$ & $8$ & $\mathbf{17.7 \pm 3.3}$ & $\mathbf{24.3 \pm 3.5}$ & $19.4\,$M  \\
\cmidrule{2-7}
 & iChan-Last & $1$ & $8$ & $\mathbf{17.7 \pm 3.3}$ & $24.2 \pm 3.5$ & $19.4\,$M  \\
\cmidrule{2-7}
 & w/o PWD & $1$ & $8$ & $17.5 \pm 3.4$ & $24.1 \pm 3.5$ & $19.4\,$M  \\
\midrule
\multirow{4}{*}{\rotatebox{90}{Baselines}} & Omni-SF & $8$ & $2$ & $12.4 \pm 2.8$ & $19.6 \pm 3.1$ & $25.7\,$M  \\
 & Omni-SF & $1$ & $8$ & $11.9 \pm 3.0$ & $19.1 \pm 3.2$ & $13.1\,$M  \\
 & PWD-SF & $1$ & $8$ & $12.8 \pm 3.0$ & $19.8 \pm 3.5$ & $13.9\,$M  \\
 & Oracle Mask & - & - & $13.8 \pm 1.8$ & $21.6 \pm 2.0$ & $16.4$K  \\
\bottomrule
\end{tabular}
\end{center}
\vspace*{-1em}
\end{table}
In Tab~\ref{tab:results}, the mean and standard deviations ($\sigma$) of the performance measures, evaluated on the test dataset, as well as the number of trainable parameters $\#$P are shown for different model configurations. The proposed triple-path transformer architecture (iChan-First with PWD) was evaluated for eight different $(N_L,R)$ configurations with $N_L\cdot R = 6$ and $N_L\cdot R = 8$. Moreover, the best-performing configuration $(N_L,R)=(1,8)$ was evaluated for a different order of the transformer blocks, viz. intra-chunk -- inter-chunk -- inter-channel (iChan-Last), as well as without the PWD encoding and decoding blocks (w/o PWD), i.e., in the Ambisonic signal domain. Finally, the baseline methods Omni-SF, PWD-SF and oracle Wiener TF masks discussed in Sec.~\ref{sec:baselines} were evaluated.

One can observe that all proposed triple-path configurations resulted in mean SI-SDRi values, at least $1\sigma$ larger than the non-oracle baselines and considerably better than the oracle mask baseline. The proposed configurations with $R=1$ performed considerably worse compared to the configurations with $R>1$. In general, the performance increased with $R$ for most cases, even though $N_L$ was decreased such that $N_L \cdot R = const$. The iChan-Last configuration performed almost identical to the corresponding iChan-First configuration. This suggests that the position of the inter-channel transformer blocks, viz. before or after the intra-chunk and inter-chunk blocks, does not significantly influence the performance, at least when the number of repeats $R$ is sufficiently large. The configuration without the PWD encoding and decoding performed only slightly worse than the corresponding proposed configuration. The small performance difference, however, suggests that there is some advantage of using the PWD encoding and decoding.

The SI-ISR measures showed a similar behavior and resulted in mean values between $22.9\,$-$24.3\,\mathrm{dB}$ for the proposed configurations, indicating that spatial distortions were low.
The non-oracle baselines yielded mean SI-ISR measures between $19.1\,$-$\,19.8\,\mathrm{dB}$, more than $1\sigma$ lower than the best-performing proposed model.

The larger triple-path transformer models with $\#\text{P} = 19.4\,$M required 27-41 GB GPU RAM during training and 3-5 hrs per training epoch on a single NVIDIA A100 GPU.
For the smaller models with $\#\text{P} = 14.6\,$M, 20-32 GB GPU RAM and 4-5 hrs per training epoch on an NVIDIA Tesla V100 GPU were required.

\section{Conclusions}
\label{sec:conclusions}
We proposed AmbiSep, a novel Ambisonic-to-Ambisonic reverberant speech separation method comprising a PWD encoder-decoder pair as well as a triple-path transformer-based deep neural network.
The proposed method and the baselines were trained and evaluated using first-order Ambisonic reverberant mixtures containing two speakers.
The best-performing configuration of the proposed method significantly outperformed the baselines, including the oracle TF mask approach.
The PWD encoder and decoder only yielded a slight mean performance increase as opposed to the same network architecture without these blocks, at least for speech separation.

\section{Acknowledgment}
The authors would like to thank the Erlangen Regional Computing Center (RRZE) for providing computing resources and support.

\bibliographystyle{IEEEbib}
\bibliography{refs}

\begin{thebibliography}{10}

\bibitem{Zotter2019}
F.~Zotter and M.~Frank,
\newblock {\em Ambisonics - A Practical 3D Audio Theory for Recording, Studio
  Production, Sound Reinforcement, and Virtual Reality},
\newblock Springer, 2019.

\bibitem{Rafaely2015}
B.~Rafaely,
\newblock {\em Fundamentals of Spherical Array Processing}, vol.~8,
\newblock Springer, 2015.

\bibitem{Jarrett2017}
D.~P. Jarrett, E.~A.~P. Habets, and P.~A. Naylor,
\newblock {\em Theory and Applications of Spherical Microphone Array
  Processing},
\newblock Springer, 2017.

\bibitem{Teutsch2008}
H.~Teutsch and W.~Kellermann,
\newblock ``Detection and localization of multiple wideband acoustic sources
  based on wavefield decomposition using spherical apertures,''
\newblock in {\em Proc. {IEEE} Intl. Conf. on Acoustics, Speech and Signal
  Processing (ICASSP)}, Mar. 2008, pp. 5276--5279.

\bibitem{Khaykin2009}
D.~Khaykin and B.~Rafaely,
\newblock ``Coherent signals direction-of-arrival estimation using a spherical
  microphone array: Frequency smoothing approach,''
\newblock in {\em Proc. {IEEE} Workshop on Applications of Signal Processing to
  Audio and Acoustics ({WASPAA})}, Oct. 2009, pp. 221--224.

\bibitem{Epain2010}
N.~Epain, C.~Jin, and A.~{van Schaik},
\newblock ``Blind source separation using independent component analysis in the
  spherical harmonic domain,''
\newblock in {\em {Intl. Symposium on Ambisonics and Spherical Acoustics}},
  Paris, France, May 2010.

\bibitem{Epain2016}
N.~Epain and C.~T. Jin,
\newblock ``Spherical harmonic signal covariance and sound field diffuseness,''
\newblock {\em {IEEE} Trans. Audio, Speech, Lang. Process.}, vol. 24, no. 10,
  pp. 1796--1807, Oct. 2016.

\bibitem{Fahim2017}
A.~Fahim, P.~N. Samarasinghe, and T.~D. Abhayapala,
\newblock ``{PSD} estimation of multiple sound sources in a reverberant room
  using a spherical microphone array,''
\newblock in {\em Proc. {IEEE} Workshop on Applications of Signal Processing to
  Audio and Acoustics ({WASPAA})}, New Paltz, NY, USA, Oct. 2017.

\bibitem{Nikunen2018}
J.~Nikunen and A.~Politis,
\newblock ``Multichannel {NMF} for source separation with {Ambisonic}
  signals,''
\newblock in {\em Proc. Intl. Workshop Acoust. Signal Enhancement ({IWAENC})},
  Tokyo, Japan, Sept. 2018.

\bibitem{Perotin2018}
L.~Perotin, R.~Serizel, E.~Vincent, and A.~Guérin,
\newblock ``Multichannel speech separation with recurrent neural networks from
  high-order {Ambisonics} recordings,''
\newblock in {\em Proc. {IEEE} Intl. Conf. on Acoustics, Speech and Signal
  Processing (ICASSP)}, Calgary, AB, Canada, Apr. 2018.

\bibitem{Perotin2019}
L.~Perotin, , R.~Serizel, E.~Vincent, and A.~Guérin,
\newblock ``{CRNN}-based multiple {DoA} estimation using acoustic intensity
  features for {Ambisonics} recordings,''
\newblock {\em {IEEE} J. Sel. Topics Signal Process.}, vol. 13, no. 1, pp.
  22--33, Feb. 2019.

\bibitem{Borrelli2018}
C.~Borrelli, A.~Canclini, F.~Antonacci, A.~Sarti, and S.~Tubaro,
\newblock ``A denoising methodology for higher order {Ambisonics} recordings,''
\newblock in {\em Proc. Intl. Workshop Acoust. Signal Enhancement ({IWAENC})},
  Tokyo, Japan, Sept. 2018.

\bibitem{Hafsati2019}
M.~Hafsati, N.~Epain, R.~Gribonval, and N.~Bertin,
\newblock ``Sound source separation in the higher order {Ambisonics} domain,''
\newblock in {\em Proc. Conf. on Digital Audio Effects}, July 2019.

\bibitem{Herzog2020}
A.~Herzog and E.~A.~P. Habets,
\newblock ``Direction and reverberation preserving noise reduction of
  {Ambisonics} signals,''
\newblock {\em {IEEE/ACM} Trans. Audio, Speech, Lang. Process.}, vol. 28, pp.
  2461--2475, Aug. 2020.

\bibitem{Ozerov2012}
A.~Ozerov, E.~Vincent, and F.~Bimbot,
\newblock ``A general flexible framework for the handling of prior information
  in audio source separation,''
\newblock {\em {IEEE} Trans. Audio, Speech, Lang. Process.}, vol. 20, no. 4,
  pp. 1118--1133, May 2012.

\bibitem{Wang2018}
D.~Wang and J.~Chen,
\newblock ``Supervised speech separation based on deep learning: An overview,''
\newblock {\em {IEEE/ACM} Trans. Audio, Speech, Lang. Process.}, vol. 26, no.
  10, pp. 1702--1726, Oct. 2018.

\bibitem{Luo2019}
Y.~Luo and N.~Mesgarani,
\newblock ``Conv-{TasNet}: Surpassing ideal time–frequency magnitude masking
  for speech separation,''
\newblock {\em {IEEE/ACM} Trans. Audio, Speech, Lang. Process.}, vol. 27, no.
  8, pp. 1255--1266, Aug. 2019.

\bibitem{Luo2020}
Y.~Luo, Z.~Chen, and T.~Yoshioka,
\newblock ``Dual-path {RNN}: Efficient long sequence modeling for time-domain
  single-channel speech separation,''
\newblock in {\em Proc. {IEEE} Intl. Conf. on Acoustics, Speech and Signal
  Processing (ICASSP)}, Barcelona, Spain, May 2020.

\bibitem{Subakan2021}
C.~Subakan, M.~Ravanelli, S.~Cornell, M.~Bronzi, and J.~Zhong,
\newblock ``Attention is all you need in speech separation,''
\newblock in {\em Proc. {IEEE} Intl. Conf. on Acoustics, Speech and Signal
  Processing (ICASSP)}, Toronto, ON, Canada, June 2021.

\bibitem{Pandey2021}
A.~Pandey, B.~Xu, A.~Kumar, J.~Donley, P.~Calamia, and D.~Wang,
\newblock ``{TPARN:} triple-path attentive recurrent network for time-domain
  multichannel speech enhancement,'' Oct. 2021,
\newblock Preprint, arXiv:2110.10757.

\bibitem{Hershey2016}
J.~R. Hershey, Z.~Chen, J.~Le Roux, and S.~Watanabe,
\newblock ``Deep clustering: Discriminative embeddings for segmentation and
  separation,''
\newblock in {\em Proc. {IEEE} Intl. Conf. on Acoustics, Speech and Signal
  Processing (ICASSP)}, Shanghai, China, May 2016.

\bibitem{Vincent2007}
E.~Vincent, H.~Sawada, P.~Bofill, S.~Makino, and J.~Rosca,
\newblock ``First stereo audio source separation evaluation campaign: data,
  algorithms and results,''
\newblock in {\em Proc. Intl. Confl. on Indep. Comp. Analysis and Signal
  Separation (ICA)}, London, UK, Sept. 2007.

\bibitem{Nachbar2011}
C.~Nachbar, F.~Zotter, E.~Deleflie, and A.~Sontacchi,
\newblock ``{AmbiX} - a suggested {Ambisonics} format,''
\newblock in {\em Ambisonics Symposium}, Lexington, KY, US, June 2011.

\bibitem{Herzog2019a}
A.~Herzog and E.~A.~P. Habets,
\newblock ``Direction-preserving {Wiener} matrix filtering for {Ambisonic}
  input-output systems,''
\newblock in {\em Proc. {IEEE} Intl. Conf. on Acoustics, Speech and Signal
  Processing (ICASSP)}, Brighton, UK, May 2019.

\bibitem{Lugasi2020}
M.~Lugasi and B.~Rafaely,
\newblock ``Speech enhancement using masking for binaural reproduction of
  {Ambisonics} signals,''
\newblock {\em {IEEE} Trans. Audio, Speech, Lang. Process.}, vol. 28, pp.
  1767--1777, May 2020.

\bibitem{Jarrett2012b}
D.~P. Jarrett, E.~A.~P. Habets, M.~R.~P. Thomas, and P.~A. Naylor,
\newblock ``Rigid sphere room impulse response simulation: algorithm and
  applications,''
\newblock {\em J. Acoust. Soc. Am.}, vol. 132, no. 3, pp. 1462--1472, Sept.
  2012.

\bibitem{Politis2017}
A.~Politis and H.~Gamper,
\newblock ``Comparing modeled and measurement-based spherical harmonic encoding
  filters for spherical microphone arrays,''
\newblock in {\em Proc. {IEEE} Workshop on Applications of Signal Processing to
  Audio and Acoustics ({WASPAA})}, New Paltz, NY, USA, Oct. 2017, pp. 224--228.

\bibitem{Kolbaek2017}
M.~Kolb{\ae}k, D.~Yu, Z.-K. Tan, and J.~Jensen,
\newblock ``Multitalker speech separation with utterance-level permutation
  invariant training of deep recurrent neural networks,''
\newblock {\em {IEEE/ACM} Trans. Audio, Speech, Lang. Process.}, vol. 25, no.
  10, pp. 1901--1913, 2017.

\end{thebibliography}

\end{document}